\documentclass[journal=aamick,manuscript=article,layout=twocolumn]{achemso}
\setkeys{acs}{articletitle = false, etalmode = truncate, maxauthors = 3}
\usepackage[version=3]{mhchem} 
\usepackage{graphicx}
\usepackage{color}
\usepackage{xfrac}
\usepackage{multirow}
\usepackage{rotating} 
\DeclareGraphicsExtensions{.bmp,.jpg,.pdf}

\usepackage{fancyhdr}

\newcommand*\mapi{\ce{CH3NH3PbI3}}
\newcommand*\cuo{\ce{Cu2O}}

\author{Jes\'us E. Castellanos-\'Aguila}
\affiliation[ugjto]{Universidad de Guanajuato, Departamento de Estudios Multidisciplinarios, Av. Yacatitas, S/N Col. Yacatitas, Yuriria, Gto. C.P. 36940 M\'exico.}
\alsoaffiliation[uchile]
{Departamento de F\'isica, Facultad de Ciencias, Universidad de Chile, Las Palmeras 3425, \~Nu\~noa 7800003, Santiago, Chile}
\altaffiliation{Contributed equally to this work}
\author{Lucas Lodeiro}
\affiliation[uchile2]
{Departamento de Qu\'imica, Facultad de Ciencias, Universidad de Chile, Las Palmeras 3425, \~Nu\~noa 7800003, Santiago, Chile}
\altaffiliation{Contributed equally to this work}
\author{Eduardo Men\'endez-Proupin}
\email{emenendez@uchile.cl}
\affiliation[uchile]
{Departamento de F\'isica, Facultad de Ciencias, Universidad de Chile, Las Palmeras 3425, \~Nu\~noa 7800003, Santiago, Chile}
\altaffiliation{Contributed equally to this work}
\author{Ana L. Montero-Alejo}
\affiliation[utem]
{Departamento de F\'isica, FCNMM, Universidad Tecnol\'ogica Metropolitana, Jos\'e Pedro Alessandri 1242, \~Nu\~noa 7800002, Santiago, Chile}
\author{Pablo Palacios}
\affiliation[upm2]{Departamento de F\'isica Aplicada a las Ingenier\'ias Aeron\'autica y Naval, ETSI Aeronáutica y del Espacio, Universidad Polit\'ecnica de Madrid, Pz. Cardenal Cisneros, 3, 28040, Madrid, Spain}
\alsoaffiliation[upm]{Instituto de Energ\'ia Solar, ETSI Telecomunicaci\'on, Universidad Polit\'ecnica de Madrid, Ciudad Universitaria, s/n, 28040, Madrid, Spain}
\author{Jos\'e C. Conesa}
\affiliation[csic]{Instituto de Cat\'alisis y Petroleoqu\'imica, Consejo Superior de Investigaciones Cient\'ificas, Marie Curie 2, 28049, Madrid, Spain}
\author{Perla Wahn\'on}
\affiliation[upm]{Instituto de Energ\'ia Solar, ETSI Telecomunicaci\'on, Universidad Polit\'ecnica de Madrid, Ciudad Universitaria, s/n, 28040, Madrid, Spain}

\title[\ce{Cu2O}/\ce{CH3NH3PbI3} interfaces]{Atomic scale model and electronic structure of \ce{Cu2O}/\ce{CH3NH3PbI3} interfaces in perovskite solar cells}

\abbreviations{MAPI, IP, EA}

\begin{document}


\thispagestyle{fancy}
\begin{abstract}
Cuprous oxide has been conceived as a potential alternative to traditional organic hole transport layers in hybrid halide perovskite-based solar cells. Device simulations predict record efficiencies using this semiconductor, but experimental results do not yet show this trend. 
More detailed knowledge about the \ce{Cu2O}/perovskite interface is mandatory to improve the photoconversion efficiency.
Using density functional theory calculations, here we study the interfaces of \ce{CH3NH3PbI3} with \ce{Cu2O} to assess their influence on device performance. Several atomistic models of these interfaces are provided for the first time, considering different compositions of the interface atomic planes. The interface electronic properties are discussed on the basis of the optimal theoretical situation, but in connection with the experimental realizations and device simulations. It is shown that the formation of vacancies in the \ce{Cu2O} terminating planes is essential to eliminate dangling bonds and trap states. The four interface models that fulfill this condition present a band alignment favorable for photovoltaic conversion. Energy of adhesion, and charge transfer across the interfaces are also studied. The termination of \ce{CH3NH3PbI3} in \ce{PbI2} atomic planes seems optimal to maximize the photoconversion efficiency. 
\end{abstract}

\section{Introduction}

Achieving high-quality interfaces is at the forefront of perovskite solar cell (PSC) research.
PSC have emerged as a strong promise for efficient and cheap solar cells, which have attained an extraordinary 25.2~\% 
record photoconversion efficiency (PCE) in less than 11 years of research\cite{kojima2009,nrelchart}. They can  also be used to boost 
the PCE of silicon solar cells, in tandem architecture, which have reached 29.1~\%  
PCE\cite{nrelchart}. PSC use hybrid organic-inorganic halides with perovskite structure (HOIHP) as light absorber. Methylammonium lead iodide (\mapi{} or MAPI) is by far the most studied material of this family. In the perovskite structure ABX$_3$, the A-, B-, and X-sites are occupied by the ions CH$_3$NH$_3^{+}$ (MA$^{+}$), Pb$^{2+}$, and I$^-$. The HOIHP family is obtained replacing iodide by other halide, replacing lead by other group IV cation, and MA by another organic cation, or cesium. The full family is composed by random alloys of the pure compounds. 
The first PSC were fabricated using MAPI as the light absorber material. However, MAPI is barely stable against decomposition into lead iodide and methylammonium iodide, causing degradation of the devices. MAPI degradation is accompanied by toxicity risks due to lead content. Therefore, much research has been done to replace lead in MAPI, and to increase stability. While lead replacement always cause a large decrease in the solar cell PCE, there are significant achievements in the stability by means of interface engineering, and by alloying the cations. 
In planar PSC, the HOIHP light absorber is sandwiched between  a p-type, and an n-type semiconductor films, known as hole transport layer (HTL), and electron transport layer (ETL), respectively. 
Sunlight absorption takes place at the HOIHP, inducing the generation of electrons and holes, which are separated at the interfaces with the ETL and the HTL, respectively.   

The ETL and HTL act as filters for the photogenerated charge carriers. Their energy bands must be properly aligned with the bands of the perovskite and the electrodes. The ETL allows the photo-excited electron flow, and block the hole flow from the perovskite by an energy barrier at the interface. For this the ETL conduction band minimum (CBM) and valence band maximum (VBM) energy must be equal or smaller than perovskite CBM and VBM energy, respectively. This is a Type II alignment\cite{yu_cardona}. The HTL allows the hole flow, and block the photo-excited electron flow by an energy barrier at the interface. For this, the HTL CBM and VBM energies must be equal or higher than perovskite CBM and VBM energy, respectively. This is also a Type II alignment. Allow us recall that a hole transfer between materials is really a transfer of an electron in the opposite direction. The alignment of energy levels is just one of the conditions required.

Other requirements are: high carrier mobility, absence of trap states, adhesion, chemical stability, non-toxicity or encapsulation, and finally costs. Some of these requirements, are contradictory and not obvious at all. For example, the use of Spiro-MeOTAD as HTL allowed the highest cell efficiencies, but it was later found to be part of the mechanism of degradation of the device\cite{wang-bisquert2019}. Recent breakthroughs rely on new HTL, such as poly(3-hexylthiophene)
(P3HT)\cite{jung_record23_2019}.

The search of ETL, and HTL, compatible with MAPI, and other HOIHP, has been guided by the empirical Anderson's rule \cite{cahenreview2019}, whereby the vacuum levels of both materials joined by the interface are aligned. Therefore, the CBM of each material is the negative of the electron affinity (EA), while the VBM is the negative of the ionization potential (IP). This has allowed to identify possible HTL\cite{Butler16_Screening_layers, wangctm2018} among p-type semiconductors with IP close to that of MAPI. 
In order that perovskite and HTL could be  joined without strain, both materials must have some atomic planes with the same in-plane lattice vectors.

The band alignment predicted by the Anderson’s rule is modified by the appearance of an interface dipole that depends on details of the interface at the atomic scale. Even for free surfaces, the ionization potential can be modified by nearly 1~eV\cite{kim2015-IPvscomposition,emara2016-IPvscomposition,quarti2017,cahenreview2019}. 
This variation is reproduced in  calculation for MAPI(001)  surfaces with different terminations\cite{quarti2017,surfacemapi}.

\cuo{}, a low cost  p-type semiconductor with relatively high hole mobility, and a nominal ionization potential of 5.37~eV, has been 
proposed\cite{cu2ohossain2015,Butler16_Screening_layers} as HTL in PSC. 
Experimental PSC with CuO, \cuo{} or CuO$_x$ as HTL have been 
reported\cite{cu2o_zuo2015, cu2o_yu2016nanoscale, cu2o_chatterjee2016, cu2o_nejand2016, cu2o_sun2016,cu2o_yu_zhikai2017,cu2o_zhang2018,cu2o_miao2019,cu2o_miao_scripta2019}, 
with a maximum PCE 
 of 17~\%,  as well as improved stability with respect to control solar cells, using spiro-MeOTAD.
 Nitrogen-doped \cuo{} has been proposed for tunnel recombination junction on tandem Si/PSC\cite{kim2018tandem}, as well as for single junction cells\cite{Han2018}.
Improved 19\% PCE has been achieved by the combination CuO$_x$/MAPbI$_{3-y}$Cl$_y$\cite{Rao2016}. 
\citet{cu2o_Liu_advsci2018} achieved a high 18.9~\% efficiency using \cuo{} nanocubes deposited from solution on a mixed perovskite  Cs$_{0.05}$FA$_{0.81}$MA$_{0.14}$PbI$_{2.55}$Br$_{0.45}$. 
Other interesting realization has been the use of \cuo{} as HTL with a buffer layer of spiro-MeOTAD\cite{ChenSciRep2018,cu2o_chen_JMCC2018,Han2018}, leading to champion PCE  over 17~\%. Reviews of the experimental achievements using copper oxides and other copper based compound HTLs can be found elsewhere\cite{review_li2016,review_bidikoudi}.
Device simulations\cite{cu2ohossain2015, WangSST2015,simul_Peltzer2017,Lin2018_simul,simul_Shasti2019,haiderJPCS2020} predict different efficiencies for MAPI solar cells using \cuo{} as HTL, the most optimistic value of which is 28~\%.

In this article, interfaces \cuo{}/MAPI parallel to the planes (001) are investigated. The effect of the stoichiometry at the interfaces atomic planes  has been studied. Atomistic models of the interfaces are presented. Their electronic properties are presented by means of projected density of states, local density of states, band alignment, charge transfer, interface states, and energy of adhesion. 
 The structure of the article is as follows. Section \ref{sec:methods} presents the methodology followed. The main results are presented Section \ref{sec:results}. The connection with the experimental solar cells is discussed in Section \ref{sec:discussion}. Section \ref{sec:conclusions} is devoted to our conclusions.

\section{Methods\label{sec:methods}}
\subsection{Electronic structure calculation}
 Electronic structure calculations
have been performed using density functional theory (DFT), as implemented in the Vienna Ab Initio Simulation Package (VASP)\cite{vasp4}. The electron-core interactions have been described by means of the projected augmented wave (PAW) method\cite{paw1,paw2}. 
Soft PAW potentials were used for O, C, and N atoms. The valence shells explicitly included in the calculations are H(1s), C(2s2p), N(2s2p), O(2s2p), Cu(3d4s), I(5s5p), Pb(6s6p). The PBE exchange-correlation functional was used. Van der Waals corrections have been applied to total energies and forces by means of the DFT--D3, or PBE--D3, method with Becke-Johnson damping\cite{dftd3grimme2011}. The wavefunctions were expanded in plane waves with energy cutoff of 400~eV.
The Brillouin Zone was sampled using a $\Gamma$ centered $2\times 2\times 1$ k-point grid. Convergence tests with $3\times 3\times 1$ were made.
The atomic coordinates and unit cell parameters have been optimized by total energy minimization.

To obtain accurate energies for the band edges,
the hybrid functional PBE0\cite{pbe0perdew,pbe0adamo} was used, including spin-orbit coupling (SOC). This method is necessary to reproduce the band gap, and the MAPI conduction band dispersion\cite{mapbi3_1}. 

The interface can be simulated by means of a supercell that contains both materials joined by interfaces, as can be seen in Fig.~\ref{fig:model1}.  
Due to periodic boundary conditions, the supercell always contains two interfaces, which should be equal or equivalent by symmetry, in order not to have a macroscopic electric field normal to the interface in the bulk materials. In some cases, depending on the crystal symmetry and orientation, it may be impossible to obtain two identical interfaces and strain-free materials. Alternatively, the supercell may contain a vacuum region, in which case there are one interface and two surfaces. 

The VBM energy $E_V$  is a bulk property of each material. For an explicit calculation of an interface between two materials,  $E_V^{(m)}$ ($m=$\cuo, MAPI) can be identified with the highest occupied energy level the wavefunction of which is delocalized over the central part of the region occupied by material $m$. The values $E_V^{(m)}$ computed in this way depend on the width of the $m$-material slab, in some cases displaying the quantum confinement effect\cite{yu_cardona,surfacemapi}. Hence,  $E_V^{(m)}$ should be computed for sufficiently wide slabs, which generally contain such a high number of atoms that prevents DFT calculations.  
However, the VBM  alignment across the interface  can be  determined by means of a two-step procedure\cite{baldereschioffset,dandrea92,dandrea93,jconesa2012tio2zno}. 
The key magnitude is the electrostatic potential energy $V(x,y,z)$, and its plane average $V(z)$ 
Assuming that the interface is parallel to the $xy$ plane, the plane-averaged electrostatic potential is defined as
\begin{equation}
\label{ec:vz}
V(z) = \frac{1}{A_{xy}} \iint V(x,y,z) dxdy   ,
\end{equation}
where $A_{xy}$ is the slab area in the $xy$ plane. 
The macroscopic electrostatic potential is defined as 
\begin{equation}
\label{ec:vmacro}
\bar{V}(z) = \frac{1}{c_1 c_2} \int\displaylimits_{-\frac{c_1}{2}}^{\frac{c_1}{2}}    \int\displaylimits_{-\frac{c_2}{2}}^{\frac{c_2}{2}}  V(z+z'+z'')  \ dz'dz'' ,
\end{equation}
where $c_1$ and $c_2$ are the periodicity lengths of $V(z)$ in each of the bulk materials. For both \cuo{} and MAPI, $c_1$ and $c_2$ have been set equal to one half of the lattice constant.
If the region occupied by each material is thick enough,   $\bar{V}(z)$ displays a plateau at the central region. The approximately constant values of $\bar{V}(z)$ at the plateaus define the values $\bar{V}^{(m)}_{slab}$ ($m=$\cuo, MAPI).

The potential energy $V(x,y,z)$ is the sum of the electrostatic electron-electron energy, and the local part of the pseudopotentials, the latter of which implies some extent of arbitrariness. However, the difference   $E_{V}^{(m)}-\bar{V}^{(m)}$ can be obtained from a bulk calculation, and it can be transferred to a slab. Therefore, the VBM at each slab is  
\begin{eqnarray}
\label{ec:vbm}
E_{V,slab}^{(m)}  &=&  \bar{V}^{(m)}_{slab} + (E_{V,bulk}^{(m)}-\bar{V}^{(m)}_{bulk}), \nonumber  \\ && (m=\mathrm{\cuo, MAPI}) .
\end{eqnarray}
The bulk calculations of $(E_{V,bulk}^{(m)}-\bar{V}^{(m)}_{bulk})$ have been made using unit cells that present the same strain as the middle region of the corresponding slab. Let us note that  VASP code sets $\bar{V}^{(m)}_{bulk}=0$.

In addition to avoiding the quantum confinement effect, the two-step process has more advantages versus a direct slab calculation of $E_{V}^{(m)}$. 
$V(z)$ and $\bar{V}(z)$ depend weakly on the exchange-correlation functional, thus it can be computed using a non expensive GGA functional. However, one must verify that the GGA error in the band edges do not cause a spurious charge transfer across the interface. 
On the other hand,  $E_{V}$  depends strongly on the functional, but it can be obtained at a high level of theory because the unit cell is a relatively small system. The \cuo{} and MAPI unit cells contain 6 and 48 atoms, respectively, while the slab supercells here considered contain 200--300 atoms. Hence, the band edges of bulk materials were calculated using the PBE0 hybrid functional.
With the PBE0 functional including SOC, the fraction of exact exchange $\alpha=0.188$ has been used for both materials. This value has been determined self-consistently\cite{marquespbe0mix,skone14pbe0a} for orthorhombic MAPI\cite{mapbi3_1}. 
The bandgaps of strain free \cuo{} and tetragonal MAPI are 2.08 and 1.77~eV, respectively, which are close to the experimental values of 2.17~eV\cite{heinemann2013} and 1.60~eV\cite{endres2016}.
 
With the same approach one can obtain the CBM energy, $E_{C,slab}^{(m)}$, as well. 
The band offsets at the interface are defined as
\begin{eqnarray}
(\Delta E_{V})_{int} & = &   E_{V,slab}^{(\mathrm{Cu}_2\mathrm{O})} - E_{V,slab}^{(\mathrm{MAPI})},  \label{ec:offset4} \\
(\Delta E_{C})_{int}&=&E_{C,slab}^{(\mathrm{Cu}_2\mathrm{O})} - E_{C,slab}^{(\mathrm{MAPI})}.
\label{ec:offset}
\end{eqnarray}
A positive sign of $(\Delta E_{C})_{int}$ means an energy barrier for the transfer of conduction electrons from MAPI to \cuo{}. For hole transfer, an energy barrier  is obtained 
when $(\Delta E_{V})_{int}$ is negative. Therefore, the ideal situation of \cuo{} as HTL is to have
  $(\Delta E_{V})_{int}\simeq 0$ , and $(\Delta E_{C})_{int}>0$. 
By construction, the bandgaps satisfy 
\begin{equation}
\label{ec:eg}
E_{g}^{(m)}=E_{C,bulk}^{(m)}-E_{V,bulk}^{(m)}=E_{C,slab}^{(m)}-E_{V,slab}^{(m)}.     
\end{equation}

In addition to the band alignments, the electronic states of the interfaces have been calculated  using the PBE0(0.188)+SOC functional. This allows to identify interface states that could act as recombination centers if their energies are within the bandgap.

\subsection{Starting models}

\begin{figure}[tbh!]
\includegraphics[width=8.3cm]{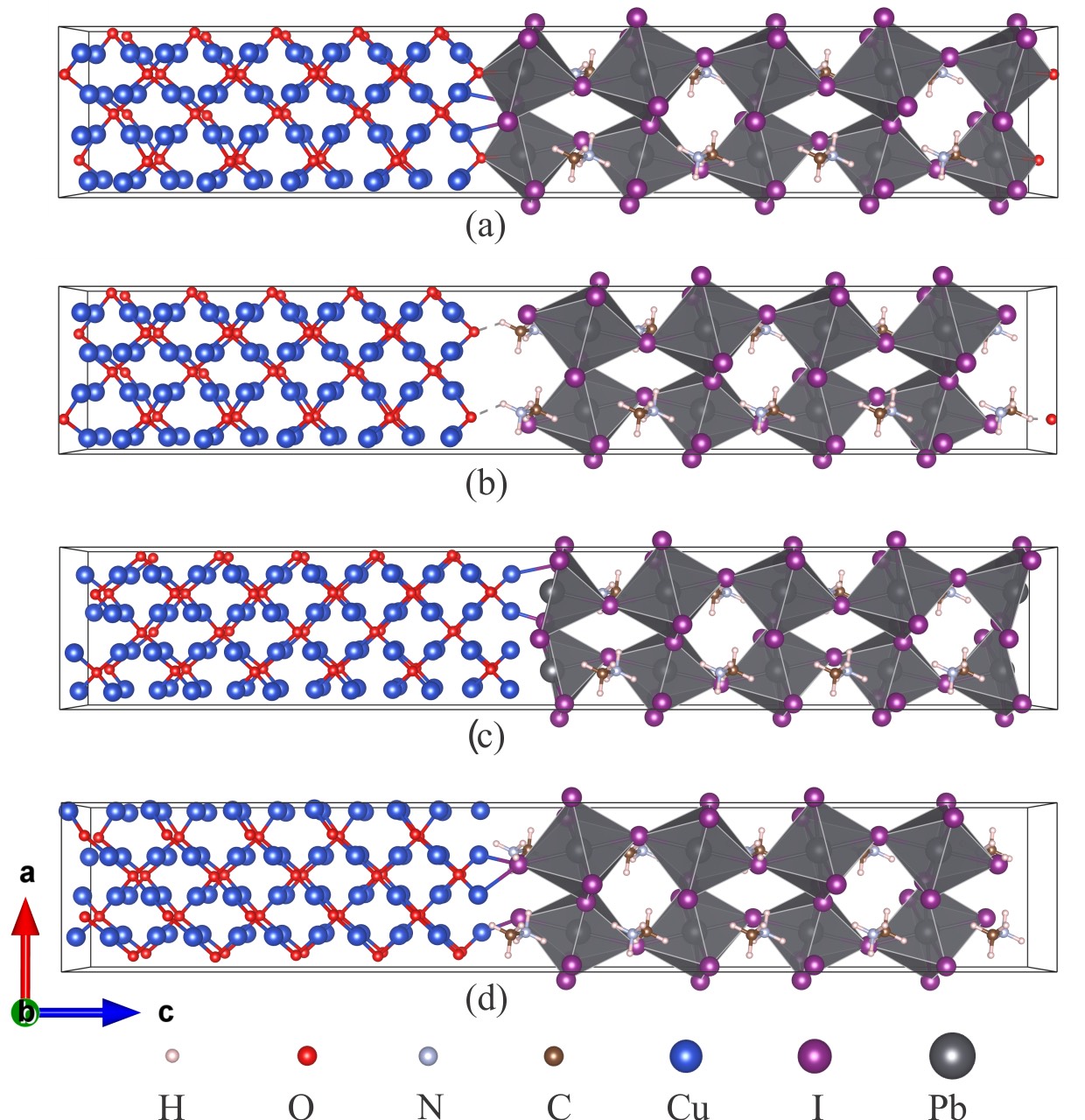}
\caption{Starting supercell models considered for the interface in (001) plane. a) O/PbI, b) O/MAI, c) Cu/PbI, d) Cu/MAI. 
Images generated with VESTA\protect\cite{vesta}. 
\label{fig:model1}}
\end{figure}

Fig.~\ref{fig:model1} shows the starting interface models for different terminations of \cuo(001) and MAPI(001). \cuo(001) surfaces are
 present on \cuo{} cubic nanoparticles \cite{cu2onano17,cu2o_Liu_advsci2018,cu2o_elseman2019}.
 \cuo(001)--X  (X=Cu, O)) denotes a surface terminated in planes of element X, while \cuo(001)--X$_{0.5}$ has half of the surface X atoms vacant. The \cuo{} surfaces without 
vacancies are Tasker III type surfaces\cite{tasker}. 
If the \cuo(001)--Cu or \cuo(001)--O 
 surfaces are neutral, there are free electrons or holes at the surface. 
 Elimination of these carriers causes the interface to have a net charge, which in turn generates an electrostatic energy, as well as conceptual and computational troubles.
 The creation of vacancies solves these problems, and turns the \cuo{} surface into Tasker II type. More complex models have been obtained for the \cuo{} surfaces in vacuum\cite{soon2007,bendavid2013,soldemo2016},
 including reconstruction and non-stoichiometric compositions. These models have not been included  in this work, in an attempt to keep the model complexity to a minimum. 
 Our interface models have been generated expanding 
 the unit cell of MAPI in tetragonal phase, and a $2\times 2\times 1$ supercell of \cuo{}, both in the [001] direction.
 The structures have been cut at selected atomic planes
 with orientation (001), and have been placed facing each other. 
For MAPI, (001) surfaces have been considered terminated either in PbI$_2$  or MAI planes, 
which are of Tasker I type. 
Hence, four interfaces has been studied in detail, as shown
in Table~\ref{tab:nomenclature} and Fig.~\ref{fig:model1}.

\begin{table}[htb!]
\caption{Interface nomenclature.}
\setlength{\tabcolsep}{0.1cm}
\label{tab:nomenclature}
\begin{tabular}{ll}
\hline\hline
Full name  & Abbrv. \\
\hline
\cuo(001)--O$_{0.5}\vert$PbI$_2$--MAPI(001) & O/PbI, \\ 
\cuo(001)--O$_{0.5}\vert$MAI--MAPI(001) & O/MAI \\ 
\cuo(001)--Cu$_{0.5}\vert$PbI$_2$--MAPI(001) & Cu/PbI \\ 
\cuo(001)--Cu$_{0.5}\vert$MAI--MAPI(001) & Cu/MAI   \\
\hline\hline
\end{tabular}
\end{table}

One problem of this method is the difficulty to accommodate two identical interfaces while keeping the continuity of bulk \cuo{} and MAPI lattices. This problem can be mitigated by variable-cell relaxation, with a careful selection of the terminating atomic planes in the starting tetragonal supercells. For the O/MAI, and O/PbI interfaces, the angles between the vectors of the relaxed supercells deviate from the right angle in less than 0.15$^{\circ}$. For the Cu/PbI interface deviations in the angles reached 0.9$^{\circ}$, but this is still manageable. For the Cu/MAI interface, the angle deviations reached 1.9$^{\circ}$,
making it difficult to define the 
reference bulk  unit cell of MAPI and \cuo{}. Therefore, for the Cu/MAI interface, a slab model with a vacuum region $\sim 15$~\AA{} in width was constructed. Thus, this model contains one single interface and two surfaces. The supercell vectors parallel to the interface were fixed at the average of the relaxed MAPI and \cuo{} lattice constants, i.e., $a=b=8.624835$~\AA{}. The relaxed vectors of the bulk phases are $a_{MAPI}=8.7161$~\AA{}, and $2a_{\cuo}=8.5336$~\AA{}. One problem of this model is the appearance of a built-in electric field inside each material, causing the macroscopic potential to have slopes instead of a plateau in the central region of each material. This is the reason why models without vacuum are preferred. For the Cu/MAI slab model with a vacuum, the built-in electric field was almost completely eliminated by adding hydrogen atoms bound to the under-coordinated oxygen atoms at the \cuo{} surface. All the atoms were allowed to relax.

\section{Results}
\label{sec:results}

\begin{figure}[tb!]
\includegraphics[width=8.3cm]{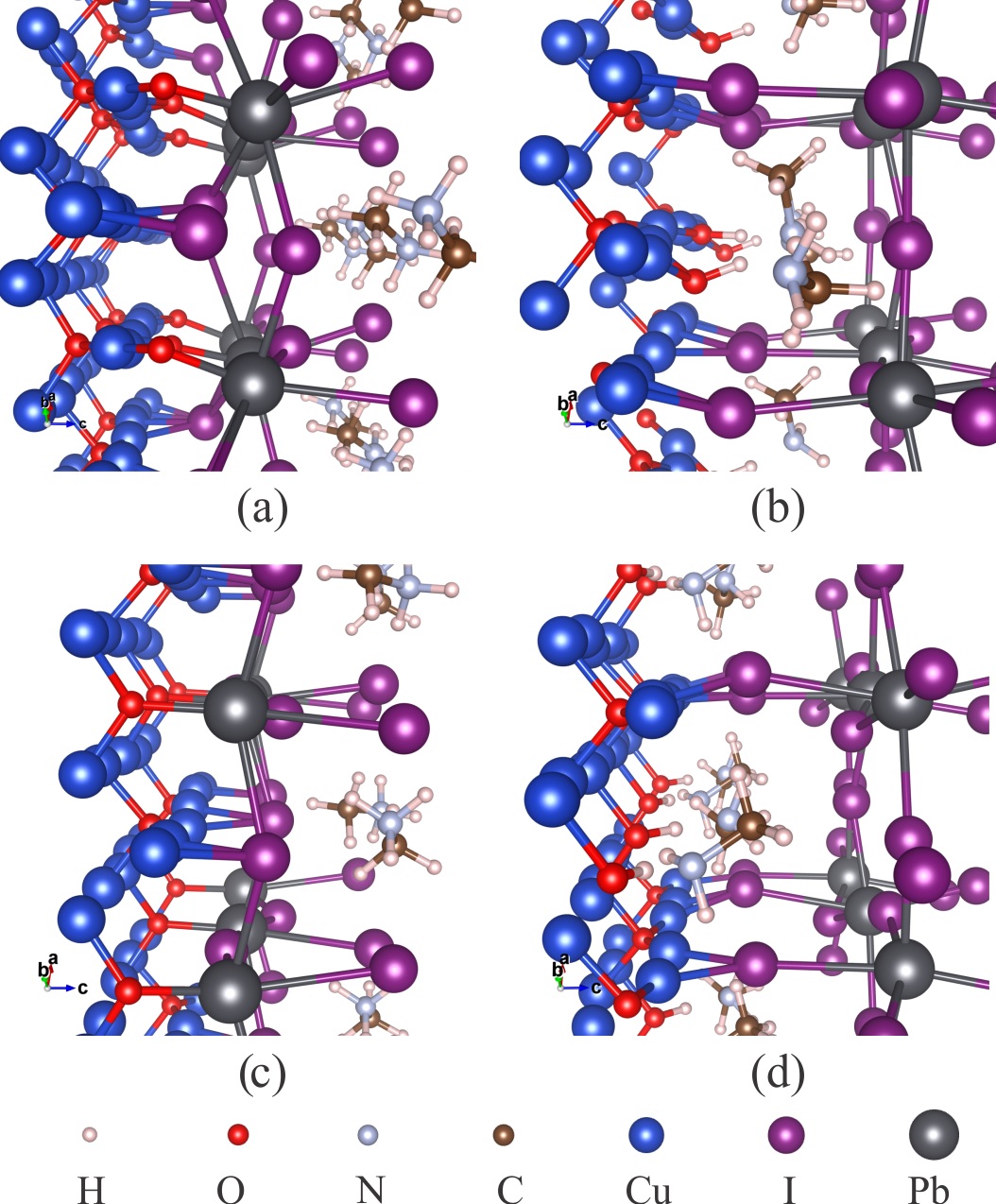}
\caption{Relaxed interface structures. a) O/PbI, b) O/MAI, c) Cu/PbI, d) Cu/MAI. The VESTA program was used to represent  and analyze the interfaces.\protect\cite{vesta}  \label{fig:interfaces}}
\end{figure}

 The optimized structures of the four interfaces proposed in this study are displayed in Fig.~\ref{fig:interfaces}. The atomic coordinates of the supercell models are given in the Electronic Supplementary Information (ESI). 
At the O/PbI and Cu/PbI interfaces (Fig.~\ref{fig:interfaces}(a) and (c), respectively), 
the O atoms are bonded with the Pb atoms, with the Pb-O bond length of $\sim$ 2.2~\AA, meanwhile, the Cu atoms are bonded with the I atoms at a bond length of $\sim$ 2.6~\AA. The strong Pb-O interactions at the interfaces pull the Pb atoms from the surface, and the Pb-I bond lengths decrease gradually as their positions vary from the surface region to the MAPI bulk region.

The four proposed interface models do not contain dangling bonds. The creation of oxygen or copper vacancies was crucial not to let interfacial atoms with missing bonds. 
\cuo{}(001) surfaces have reactive oxygens, either due to termination in oxygen planes, or due to the copper 
vacancies in Cu-terminated surfaces. In all models, the reactive oxygens either bind to lead atoms of Pb-terminated MAPI, 
or form hydroxyls capturing hydrogens from MAI-terminated MAPI. In the latest case, the methylammonia undergo dissociation and the remaining amine groups keep close to the hydroxyls, fulfilling the hydrogen bonding criteria. An alternate model of the Cu/MAI interface with one single OH and one non-dissociated MA$^{+}$ was obtained, which has a higher total energy by 0.5~eV. 
Copper atoms find ionic partners in the iodine atoms of PbI$_2$ or MAI surface layers. 
On the other hand, for PbI-terminated MAPI, the broken Pb-centered octahedral are completed by the interfacial oxygens. 
For MAI-terminated MAPI, the broken I-Pb bonds are replaced by ionic bond of iodines with two coppers.
For Cu/PbI and O/PbI interfaces, there are no changes associated with the methylammonium groups.

For the sake of evaluating the stability of the structure, the interface energy of adhesion was calculated and summarized in Table~\ref{tab:formation}. 
The energy of adhesion has been calculated as 
\begin{equation}
E_{A}=\frac{1}{N_i}\left(E_{\cuo/\textrm{MAPI}}-E_{\cuo}-E_{\mathrm{MAPI}}\right)    
\end{equation}
where $E_{\cuo/\textrm{MAPI}}$, $E_{\cuo}$, and $E_{\mathrm{MAPI}}$ are the energies of the three supercells than contain both materials, only the \cuo{} part, and only the MAPI part, respectively. $N_i$ is the number of interfaces present in the supercell, i.e. $N_i=1$ for the Cu/MAI model 
with vacuum, and $N_i=2$ for the other models. 
The energies of adhesion of the four interfaces are all negative, and are large enough to consider that the four \cuo{}/MAPI interfaces are stable. The Cu/PbI interface displays the stronger adhesion ($E_A=-1.29$~J/m$^{2}$). 
The \cuo{} and MAPI strain energy densities are provided as well. The latter have been calculated for unit  cells  with three dimensional periodic conditions, with the strained lattice vectors as in the interface supercell. The strain energy arises from the small lattice mismatch between \cuo{} and MAPI. These values are provided as indication of the degree of deformation that these materials must undergo to couple in lattice-matched interfaces. The precise values 
depend on the amount of each material that is present in the simulation supercells. Experimentally, this energy depends on grain depth, and can be relaxed by the formation of grain boundaries.
 
\begin{table}[tbh!]
\caption{Energy of adhesion $E_A$ and strain energy density $E_S$ for the different interfaces. 
\label{tab:formation}}
\begin{tabular}{lccc}
\hline\hline
System &\multicolumn{1}{c}{ $E_{A}$ (J/m$^2$)} &  \multicolumn{2}{r}{$E_S$ ($10^{7}$~J/m$^3$)}  \\
       &        & \cuo{} & MAPI\\  
       \hline 
O/PbI  & -1.19	& 0.29 & 11.2   \\
O/MAI  & -1.01	& 0.54  & 8.6   \\
Cu/PbI & -1.29  & 2.1 & 9.1  \\
Cu/MAI & -1.12  & 0.52 & 6.5   \\
\hline\hline
\end{tabular}
\end{table}

The results of the band alignment computed by means of 
Eq.~(\ref{ec:offset4}) and (\ref{ec:offset}) are 
summarized in Table~\ref{tab:electronic2}.  The plane averaged and macroscopic electrostatic potential for the 
\cuo{}/MAPI interfaces are shown in Fig.~S1 of the ESI.
For all interfaces, the calculated $\Delta E_{V}$ and $\Delta E_{C}$ are in the range of  
-0.08--+0.88 eV  and +0.16--+1.11 eV, respectively.  Positive values (Type II alignment\cite{yu_cardona})
allows to separate the electrons and holes at different sides of the interface, which is useful for the interface HTL/absorber in solar cells. However,  small negative valence band offset are still useful, as will be discussed later. 

Table ~\ref{tab:electronic2} contains the bandgaps obtained for bulk MAPI and \cuo{}, in the distorted configurations obtained from the centers  of the slabs, using the PBE0(0.188)+SOC functional.   
The  variation in the bandgap is caused  by the strain induced to accommodate the latticed mismatch and bond matching across the interface. The strain has been 
simulated by defining the bulk lattice vectors from interatomic distances at the slab centers.  
These bandgap variations are small enough not to invalidate the results, and can be easily assimilated as local fluctuations.

To evaluate the error due to a particular choice of functional, Table S2 shows the values obtained using the HSE functional\cite{hse06}, including the SOC.  
The results are similar, although one can appreciate slightly smaller band offsets with the HSE+SOC functional. The difference is explained by 
the fact that the different fractions of exact exchange were used for each material, tuned to reproduce the bulk bandgaps. Thus, with the HSE+SOC calculation, the MAPI VBM is somewhat higher, while the \cuo{} VBM is lower.  However, the differences are not larger than other sources of error in our computational setup, such as the 
supercell-dependent strain above discussed. Therefore, the results with PBE0(0.188)+SOC and HSE+SOC are qualitatively equivalent.

\begin{table}[bth!]
\setlength{\tabcolsep}{0.1cm}
\renewcommand{\arraystretch}{1.3}
\caption{Electrostatic potentials at slab centers, valence band maxima,  bandgaps at both sides of the interfaces, and band offsets. All values in the Table are in eV.
\label{tab:electronic2}}
\begin{tabular}{lcccc}
\hline\hline
Interface & O/PbI & O/MAI & Cu/PbI & Cu/MAI \\ 
              \hline
$\bar{V}^{(\mathrm{\cuo})}_{slab}$   & -1.67 & -1.22	& -1.24 & -2.96  \\   
$\bar{V}^{(\mathrm{MAPI})}_{slab}$   & 1.10 & 0.65 	& 1.31 & -0.49 \\
$E_{V,bulk}^{(\cuo)}$	            & 3.97   & 3.95	& 3.99 & 3.91 \\
$E_{V,bulk}^{(\mathrm{MAPI})}$	    & 1.28 	& 1.20 	& 1.18 & 0.99 \\
$E_{g}^{(\cuo)}$					    & 2.07	& 2.05	& 2.01 & 2.04 \\
$E_{g}^{(\mathrm{MAPI})}$	         	& 1.83 	& 1.82 	& 1.86 & 1.89 \\
$E_{V,slab}^{(\cuo)}$	            & 2.30  & 2.73	& 2.75 & 0.95 \\
$E_{V,slab}^{(\mathrm{MAPI})}$	    & 2.38 	& 1.85 	& 2.49 & 0.51 \\
$E_{C,slab}^{(\cuo)}$	            & 4.37  & 4.79	& 4.76 & 2.99 \\
$E_{C,slab}^{(\mathrm{MAPI})}$	    & 4.21 	& 3.67 	& 4.34 & 2.40 \\
$(\Delta E_{V})_{int}$          	& -0.08	& 0.88 	& 0.26 & 0.45 \\
$(\Delta E_{C})_{int}$ 	            & 0.16 	& 1.11 	& 0.42  & 0.59 \\
\hline\hline
\end{tabular}
\end{table}

\begin{table}[bth!]
\setlength{\tabcolsep}{0.1cm}
\renewcommand{\arraystretch}{1.3}
\caption{Band edges, bandgaps, and band offsets obtained  from the self-consistent slab calculation with PBE0(0.188)+SOC. All values in the Table are in eV. $E_{g,slab}^{(m)}=E_{C,slab}^{(m)}-E_{V,slab}^{(m)}$.
\label{tab:electronic4}}
\begin{tabular}{lcccc}
\hline\hline
Interface & O/PbI & O/MAI & Cu/PbI & Cu/MAI \\ 
              \hline
$E_{V,slab}^{(\cuo)}$	            & 2.29  & 2.66	& 2.66 & 0.88 \\
$E_{V,slab}^{(\mathrm{MAPI})}$	    & 2.38 	& 1.74 	& 2.44 & 0.36 \\
$E_{C,slab}^{(\cuo)}$	            & 4.50  & 4.92	& 4.89 & 3.09 \\
$E_{C,slab}^{(\mathrm{MAPI})}$	    & 4.41 	& 3.72 	& 4.47 & 2.49 \\
$E_{g,slab}^{(\cuo)}$				    & 2.21	& 2.25	& 2.23 & 2.21 \\
$E_{g,slab}^{(\mathrm{MAPI})}$	    	& 2.03 	& 1.98 	& 2.03 & 2.13 \\
$(\Delta E_{V})_{int}$   		& -0.09	& 0.92 	& 0.22 & 0.52 \\
$(\Delta E_{C})_{int}$ 	    	& 0.09 	& 1.19 	& 0.42 & 0.60 \\
\hline\hline
\end{tabular}
\end{table}

\begin{figure}[tb!]
\includegraphics[width=8.5cm]{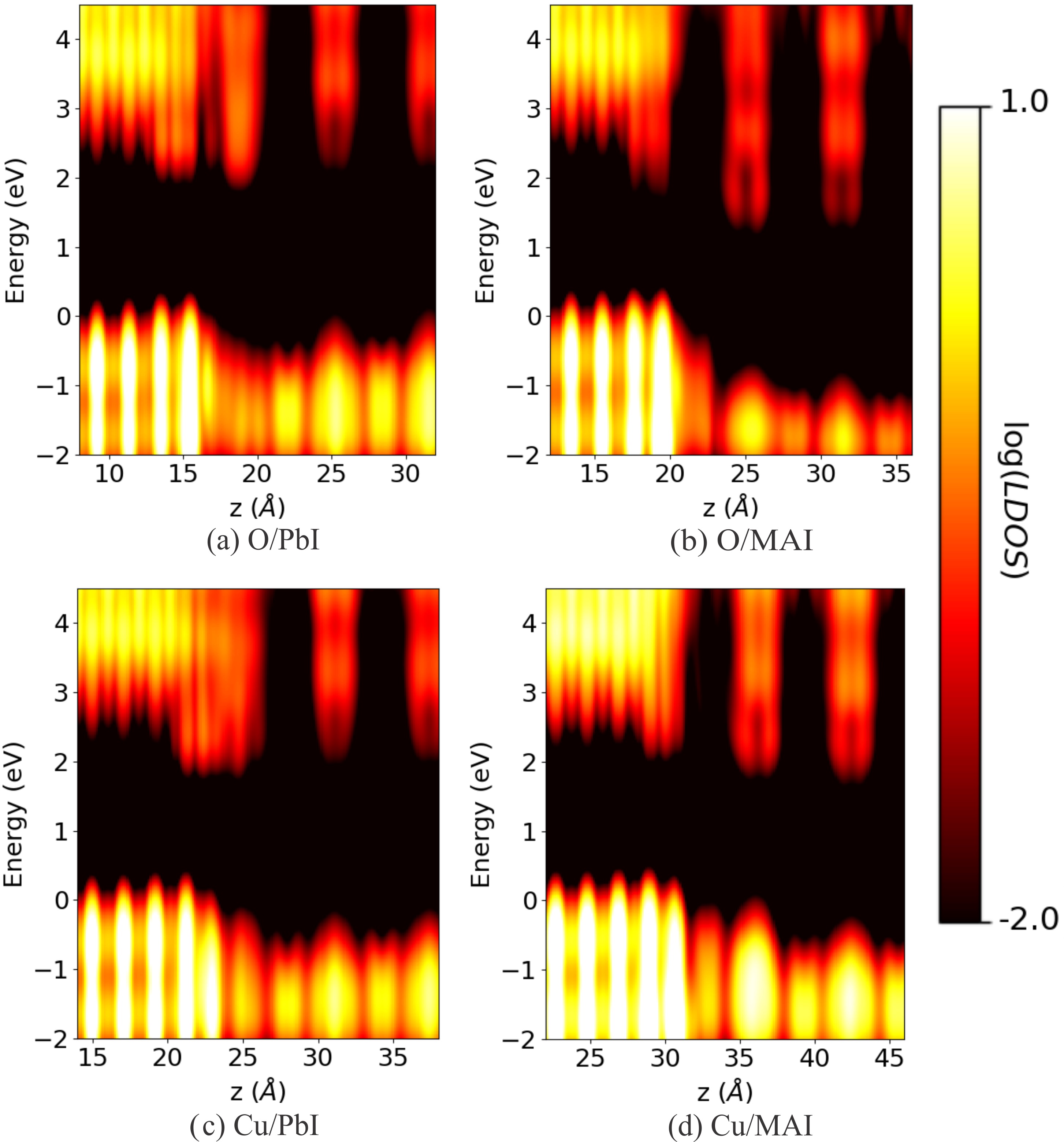}
\caption{Logarithm of LDOS $\times V_{cell}$ averaged in the $xy$ plane calculated between the MAPI and the \cuo{}     interfaces. \label{fig:LDOS}}
\end{figure}

To investigate the existence of electronic states localized at the interfaces, 
we have made a calculation of the supercell with the hybrid functional PBE0(0.188)+SOC. 
The energies and isosurfaces of the frontier orbitals (squared), at $\mathbf(k)=0$, are shown in Fig. S2-S5. 
According to their localization, these orbitals can be classified as localized at \cuo, MAPI, or at the interfaces. 
The highest and lowest energies of the occupied and unoccupied states, respectively, localized at either \cuo{} or MAPI, define the respective valence and conduction band edges ($E_{V,slab}^{(\cuo)}$, etc).
Their squared wavefunction isosurfaces can be identified among the set of orbitals represented in Fig. S2-S5. 
The band edges, bandgaps,
and band alignment inferred from these numbers are summarized in Table~\ref{tab:electronic4}. 
Due to the small thickness of the slabs that can be computed, the bandgaps and the individual band edges are modified by the quantum confinement effect.  
The band edges in Table~\ref{tab:electronic2} are free of confinement effects, hence the band alignment of Table~\ref{tab:electronic2} are better estimators, in principle. 
From Fig.~S2 and S4, one can see that the lower unoccupied molecular orbitals (LUMO) 
in O/PbI and Cu/PbI are interface states.
Fig.~S4 shows that the two interfaces of the Cu/PbI model, have localized states with different energies, revealing some asymmetry in the achieved model. This asymmetry is evident in Fig.~S6 (see below).
The in-gap interface states are dangerous for the photovoltaic conversion, they can trap photoexcited electrons and holes, facilitating recombination\cite{kobayashi,street}. 
The energy and space dependence of the set of single-particle eigenstates can be combined in the local density of states (LDOS)\cite{tomasldos}. Fig.~\ref{fig:LDOS} shows the LDOS at the interface region; full LDOS data are shown in Fig.~S6. An energy smearing of 0.27~eV was used to compute the LDOS.
The LDOS display the band alignment according to the values of Table~\ref{tab:electronic4}, and the interface states.
Fig.~\ref{fig:LDOS}(b and d) suggest that the O/MAI and Cu/MAI interface also present interface states with energies near the CBM. However, these interface states are higher in energy than the MAPI CBM (are not in-gap  interface states), allowing electrons in these states to return to MAPI.
Therefore, holes can transferred from MAPI to \cuo{} without being trapped, although they can be attracted and recombine with electrons trapped at O/PbI and Cu/PbI interfaces. 

\begin{figure}[tb!]
\includegraphics[width=8.5cm]{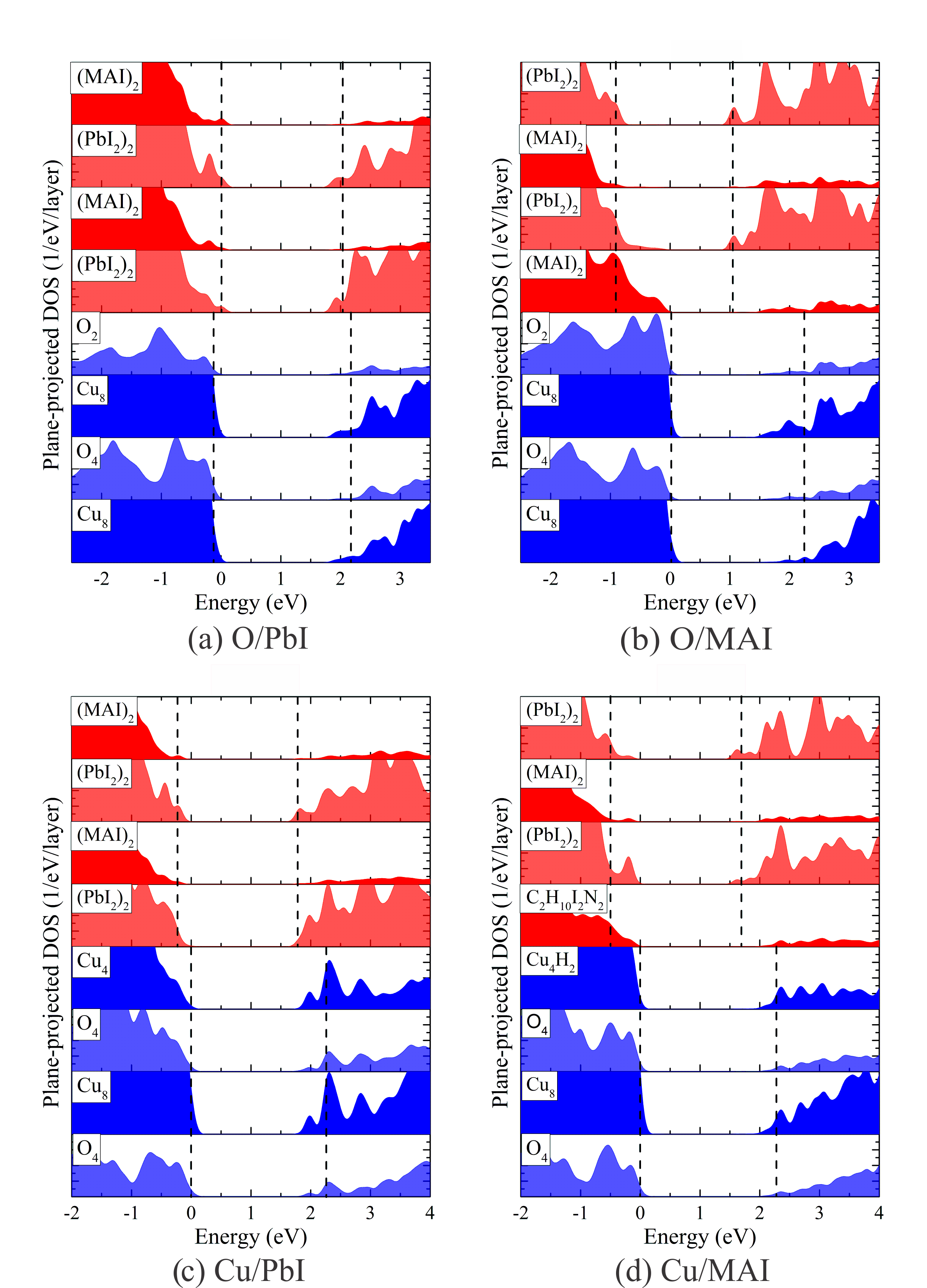}
\caption{PDOS projected on the atomic planes near each interface proposed in this study. The dotted vertical lines corresponds to the valence band maximum and the conduction band minimum for each material.  \label{fig:PDOS}}
\end{figure}

The LDOS at MAPI near the band edges is much smaller than the LDOS at \cuo{}, causing  
the former not to be discernible in the Fig.~\ref{fig:LDOS}. The band edges are better shown  
by the site projected densities of states (PDOS).  
The atomic PDOS have been added up by atomic layers
parallel to the interfaces, and are shown in Fig.~\ref{fig:PDOS} 
and Fig. S2-S5. There, one can appreciate the small PDOS at MAPI near the band edges. 

In Fig.~\ref{fig:LDOS} and \ref{fig:PDOS} the energy zero has been set at the higher occupied molecular orbital (HOMO) energy. For the Cu/MAI model, the HOMO-1 energy has been used, as the HOMO is a surface state with no interest for our study. 
For every interface model, Fig.~\ref{fig:PDOS} shows the PDOS (from the PBE0(0.188)+SOC calculation) for the interface and the first three sub-interface layers. It can be observed that the bandgap varies non-monotonically across the interface due to the local variation of the band edges. 
For the O/PbI  interface (Fig.~\ref{fig:PDOS}(a)), the valence and conduction manifold near the Fermi energy is almost completely occupied by states lying on the PbI layers.
However, for the O/MAI, Cu/PbI, and Cu/MAI interfaces Fig.~\ref{fig:PDOS}(b-d)), this behavior is changed. The larger contribution to the valence manifold near the Fermi energy is due to the \cuo{} states, while the conduction manifold is conformed of MAPI states. Such behaviors results from  complex rehybridizations
of the valence orbitals upon formation across the interface and from changes in the local potential due to the deviation from the bulk environment at the interface, which lead to the interface electron depletion or accumulation. Also, the proton transfer in the O/MAI and Cu/MAI
favors the stabilization of the interface oxygen orbitals. 
Besides, the projection of interface states over the interfacial layers is present for the O/PbI model as expected , but there is no visible projection for the Cu/PbI model. 
As mentioned above, the Cu/PbI model presents two different interfaces states (as shown in Fig.~S4 and S6), the one with in-gap energy is not localized at the layers corresponding to the PDOS of Fig.~\ref{fig:PDOS}.
In contrast with the conduction band interface states, there are not in-gap valence band interface states for any model, as shown in Fig.~\ref{fig:LDOS}, \ref{fig:PDOS}, and S1-S4.

\begin{figure*}[t!]
\centering
\includegraphics[width=0.9\linewidth]{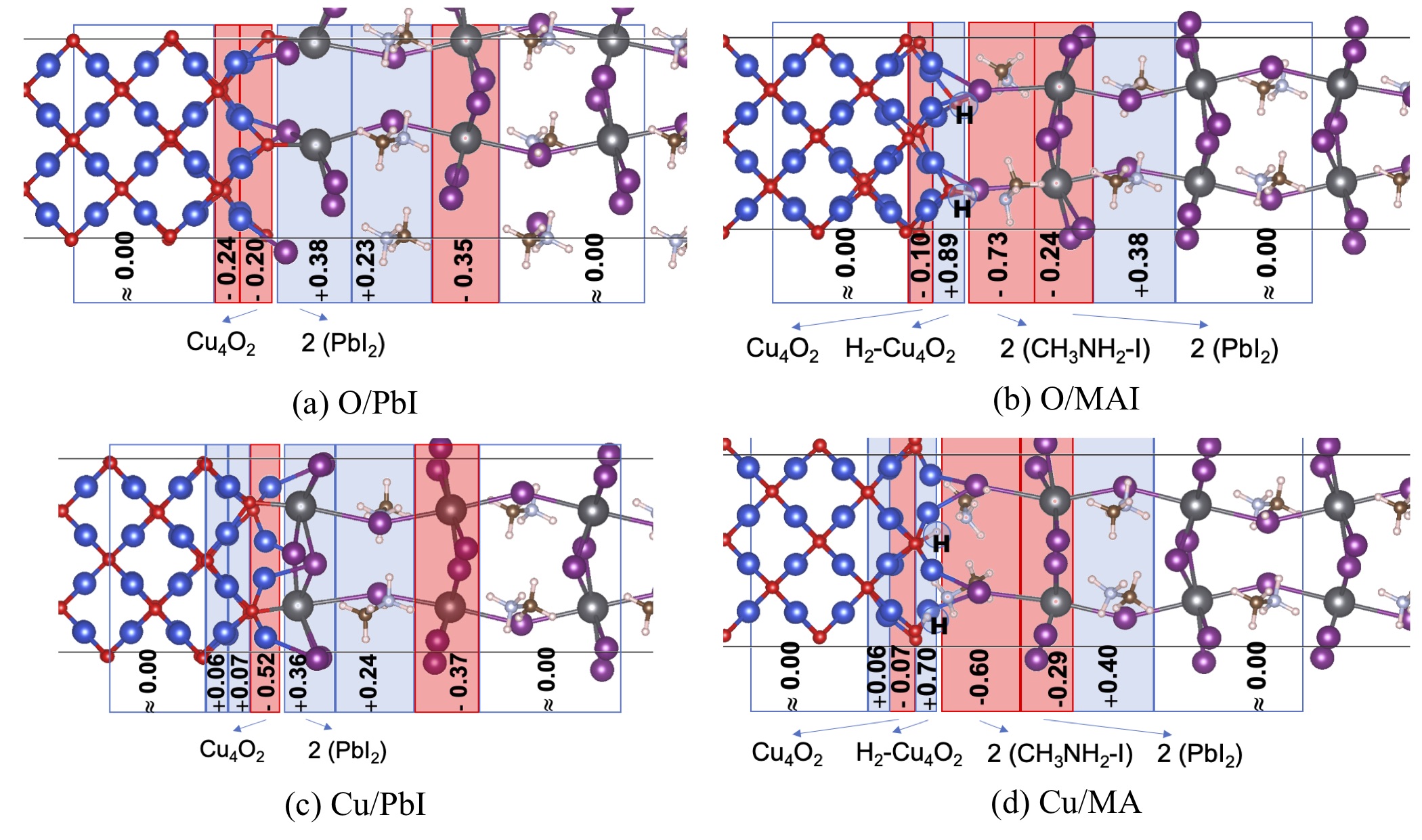} 
\caption{Bader charges summed by layers for the four interface models.}
\label{fig:bader}
\end{figure*}

A word of caution is in order, concerning the above discussed trap states in O/PbI and Cu/PbI models. The quantum confinement energy shift is present on the VBM and CBM states, mainly in $E_C^{(MAPI)}$. For a MAPI-in-vacuum slab that has the same thickness as our models\cite{surfacemapi}, the confinement energy shift is 0.24~eV. A MAPI slab free of quantum confinement would be at least three times thicker, which is computationally not  affordable. The subtraction of 0.24~eV from $E_C^{(MAPI)}$ (of Table~\ref{tab:electronic4}) lowers the CBM to 4.17 and 4.23~eV, for O/PbI and Cu/PbI models respectively. Embedding MAPI in \cuo{} may alter the quantum confinement shift. Because of this, an independent estimation is obtained comparing the $E_C^{(MAPI)}$ values of Table~\ref{tab:electronic2} and Table~\ref{tab:electronic4}. With both procedures, $E_C^{(MAPI)}$ has lower energy than the interface state of O/PbI, but it is still higher than interface state in the Cu/PbI interface. In brief, the in-gap interface states of O/PbI in Table~\ref{tab:electronic4} are likely to be artifacts of a thin model. For Cu/PbI, one of the interface states cannot be removed from the gap with a thicker model. However, the energy difference is small, and it can be blurred by the thermal motion\cite{surfacemapi,monteromd2016}. 

To analyze the spatial charge redistribution across the \cuo{}/MAPI interfaces, we have computed the 
 Bader atom charges\cite{bader90} using Henkelman's Group program\cite{Tang09}. The Bader atom charge 
is the total (including the core) electric charge within the atom boundary, which is established by the topological properties of the electron density. Fig.~\ref{fig:bader} shows the Bader charges summed by layer near the interfaces.  In the partition of \cuo{} in layers, 
the 2:1 stoichiometry has been respected. For layer boundaries coincident with Cu or O planes, as in Fig.~\ref{fig:bader}, the Bader charges of the boundary atoms are equally distributed to each layer on both sides of the boundary.
For the O/MAI and Cu/MAI interfaces, the Bader charges of the two hydrogens of OH groups are formally included in the Cu$_4$O$_2$ layer in Fig.~\ref{fig:bader}, to assess how charges are organized at the interface after dissociation or deprotonation of MA$^{+}$.
According to the above described partition, each interface formation involves a net electron transfer from the  MAPI to \cuo{}, even in the case of MAI-type interfaces where two protons (per interface) are also transferred, with $+0.64$ Bader hydrogen charge. This electron transfer is localized at very few layers near the interfaces. The inside of \cuo{} and MAPI remain neutral. The inner layers of the MAPI slab are neutralized when considering the charge of a stoichiometric MAPI unit: a PbI$_2$ layer and a MAI layer (see Fig.~S11).

\begin{figure}[b!]
\includegraphics[width=8.3cm]{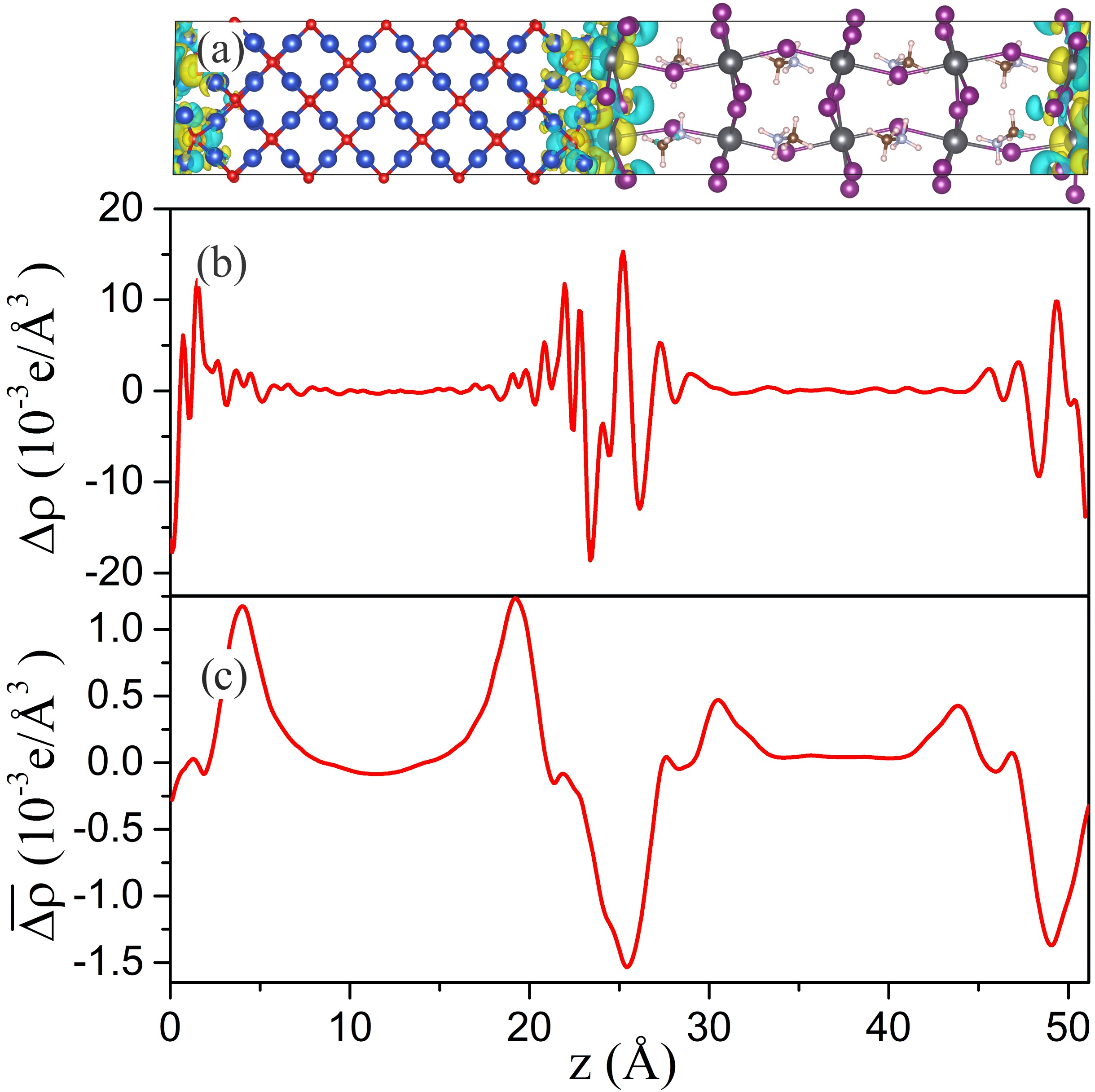}
\caption{(a) Electron density difference, and the corresponding planar (b) and macroscopic (c) averages at the Cu/PbI interface. The yellow region represents charge accumulation, and the cyan region indicates charge depletion. The isosurface value is 0.002028~$e$/\AA$^{3}$. \label{fig:CDD}}
\end{figure}

The charge separation described for the formed interfaces suggests the formation of interface dipoles. However, the way of partitioning the charges between the bulk and the surface regions is not unique, and this selection can significantly affect the magnitude of the estimated dipole. Instead, we perform a qualitative analysis with the charge by layers scheme. Fig.~\ref{fig:bader} shows that dipoles are oriented in opposite directions depending on the MAPI surface composition and their particular electron and proton transfer.

It is not obvious that Bader charges describe electron charge transfer, rather than the adjustment of the atom boundaries. As complement to the previous analysis, the electron charge density difference $\Delta\rho(x,y,z)$ has been computed by subtracting from the electron density of a \cuo/MAPI heterostructure, the densities of the only-\cuo{}, and only-MAPI slabs. In the last two, the atoms were not relaxed.  
An isosurface representation of $\Delta\rho(x,y,z)$  can be seen in Fig.~\ref{fig:CDD}(a), and 
S6-S8. The electron density accumulation is shown as the yellow isosurface, while the electron depletion is shown as the cyan isosurface. There are significant differences in the interfacial bonding characteristics 
of the two sides. First, the charge redistribution mostly takes place at the \cuo{}/MAPI interface region, 
in agreement with the picture of Bader charges. 
Second, electron charge depletion happens mainly in the interfacial regions of the PbI$_2$ layer. This signals a decrease in lateral Pb-I bond length in favor of forming new bonding in the interfacial regions, while the lateral Cu-Cu interatomic distance in the \cuo{} slab increases to match with the MAPI interface (for  more detail see Fig.~S10).
This is related to the changes in the average interlayer distance shown in Table S1. There is a reduction in the distance between layers in \cuo{} and a slight increase in these distances in MAPI.
Furthermore, though the structure has an amount of electron charge accumulation along the O-Pb and Cu-I direction, the O-Pb bond presents a roughly spherical hump (See Fig.~S10), which is slightly closer to the O atom than to the Pb atom. This is the characteristic of a highly polar covalent bond. 

The planar-averaged electron density difference $\Delta\rho(z)$ is shown in Fig.~\ref{fig:CDD}(b), and the macroscopic average, in the sense of Eq.~(\ref{ec:vmacro}) is shown Fig.~\ref{fig:CDD}(c). As can be  seen, the positive values represent electron accumulation and negative values indicate electron depletion. 
The change of electron charge density at the Cu/PbI interface shows that electrons are transferred from the MAPI to the \cuo{} through the interface, in agreement with the Bader charge picture. In general, when the equilibrium state is reached, this electron charge transfer leads to the formation of a built-in electric dipole at the Cu/PbI interface. 
Similar results are found for the other interfaces and are shown in the ESI (Fig. S7-S9).

As a summary of charge effects, the interface present electron charge transfer from MAPI to \cuo{} independent of interface composition, due to inductive effect. Also particular charge effects differentiate PbI$_2$ and MAI interfaces, which explain the interface dipole inversion in different cases. The former interface presents delocalization of the dangled bond state in Pb along the interface and inward \cuo{}, this is a pure electron charge transfer. The later presents a proton migration (acid-base reaction) which as well produces an electronic migration by tidal effect. 

\begin{figure}[ht!]
\includegraphics[width=8.3cm]{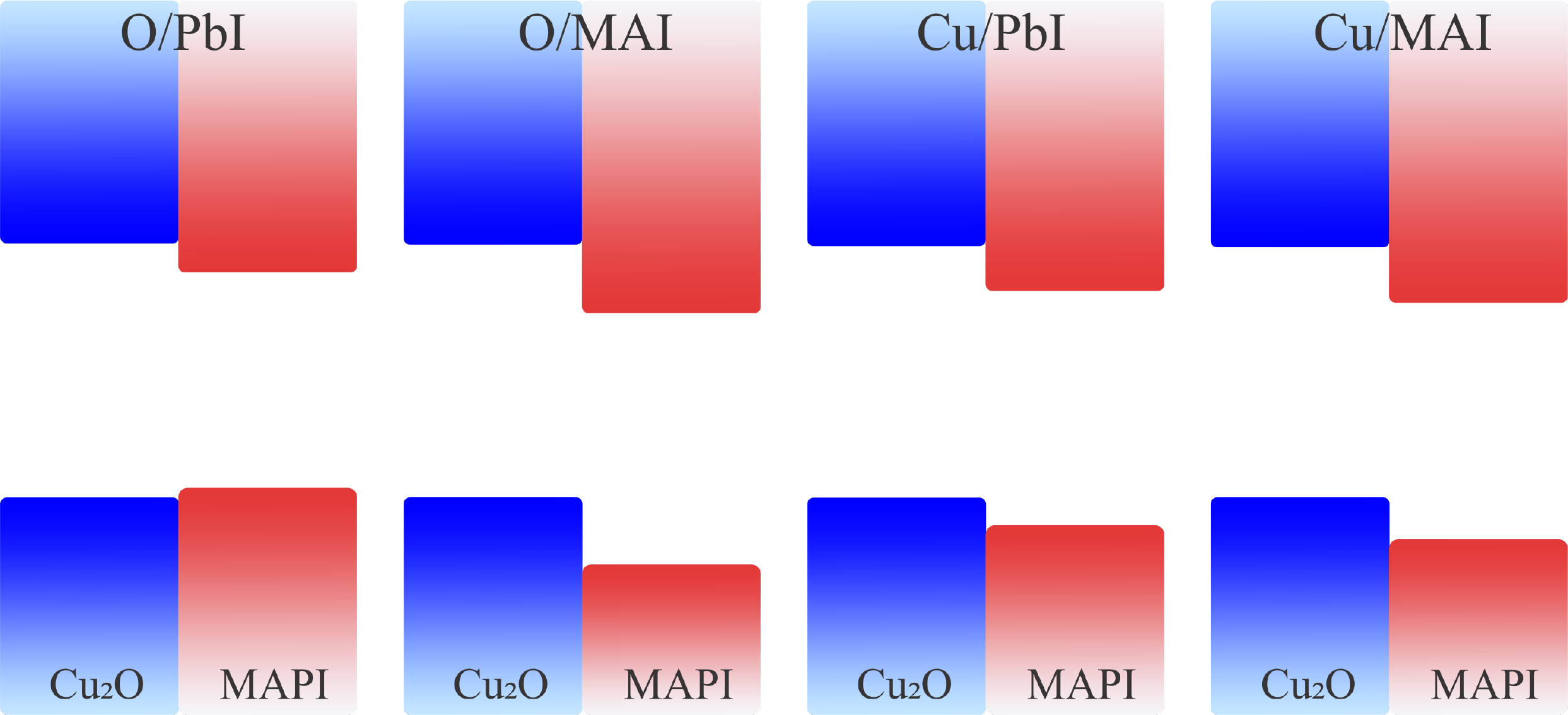}
\caption{Predicted band alignment between the MAPI and the \cuo. The band offsets for each interface, corresponds to the calculated in Table ~\ref{tab:electronic2}.
\label{fig:alignment}}
\end{figure}

Finally, a constructed theoretical energy-level band-alignment diagram based on the averaged electrostatic potential  for the \cuo/MAPI Interfaces are illustrated in Fig.~\ref{fig:alignment}. According to our results, when light is irradiated in a \cuo{}/MAPI interface, electrons are excited to the conduction band on the MAPI side of the interface; simultaneously, photoexcited holes are left in their valence band. The holes in the valence band of the MAPI side transfer to the valence band on the \cuo{} side, since the latter VBM is higher. 
The negative band offset found for the O/PbI interface seems unfavorable for the 
hole transfer, but it does not necessarily decreases solar cell performance, as it will be 
discussed below.

\section{Discussion}
\label{sec:discussion}

\begin{figure}[b!]
    \includegraphics[width=8.3cm]{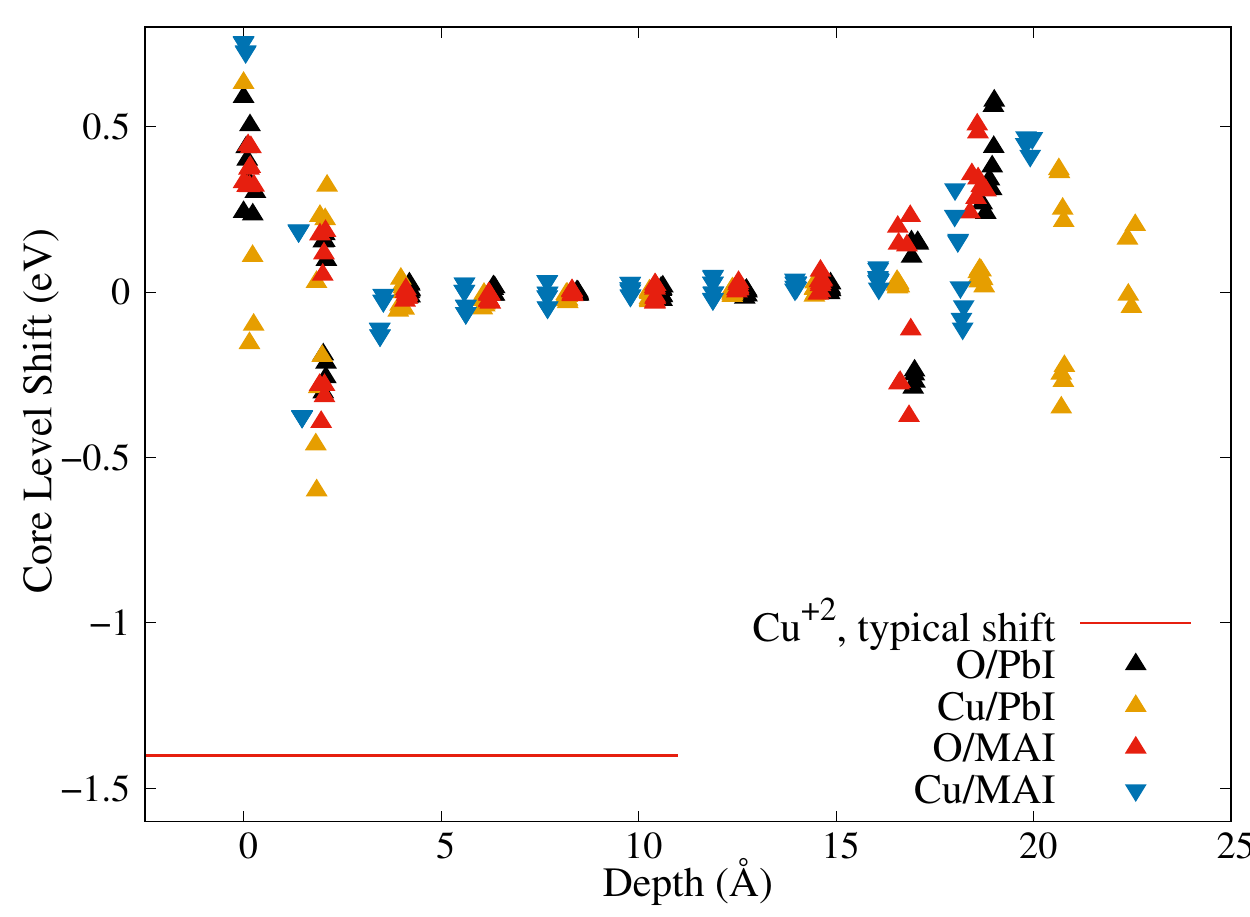}
    \caption{Copper core level shifts, respect to Cu$^{+}$ core level binding energies (middle Cu$_{2}$O copper atoms), along the interface depth coordinate.}
    \label{fig:cls}
\end{figure}

In this section, our results are examined in  connection with  experimental solar cells. 
First, most of the best performing PSC that use copper oxide as HTL have been prepared 
growing the perovskite over a copper oxide ultrathin films named CuO$_x$ rather than \cuo{}. 
To the best of our knowledge, the highest PCE of this group (19\%)  has been achieved by the combination CuO$_x$/MAPb(I$_{1-y}$Cl$_y$)$_3$\cite{Rao2016}. Devices using pure MAPI have achieved 17\% PCE\cite{cu2o_sun2016,cu2o_yu_zhikai2017}. The performance improvement of MAPb(I$_{1-y}$Cl$_y$)$_3$ was traced to an increased perovskite grain size and  carrier lifetime\cite{Rao2016}, although 
the incorporation of Cl into the \mapi{} lattice is minimal.
No crystallographic characterization of CuO$_x$ films has been 
provided, presumably due to its small thickness.  X-ray photoelectron spectroscopy (XPS) shows a variable relative amount of Cu$^{+}$ and Cu$^{2+}$ species on the exposed side of CuO$_x$\cite{Rao2016,cu2o_yu2016nanoscale,cu2o_yu_zhikai2017,cu2o_zhang2018,cu2o_elseman2019,cu2o_miao2019,cu2o_miao_scripta2019}. 
Cu$^{+}$ and Cu$^{2+}$ are  typically associated with \cuo{} and CuO, respectively. However, 
there is another crystalline phase, paramelaconite (Cu$_4$O$_3$)\cite{heinemann2013} 
which has equal amounts of  Cu$^{+}$ and Cu$^{2+}$. A disordered paramelaconite phase exists\cite{datta78} 
 with approximate stoichiometry Cu$_{64}$O$_{57}$ having  78\%  Cu$^{2+}$, and 22\% Cu$^{+}$, close to the ratio reported in Ref.~\citenum{cu2o_yu_zhikai2017}. 
  The bandgaps of CuO and Cu$_4$O$_3$, being smaller than 1.7~eV\cite{heinemann2013},  are too low to to act as HTL while blocking the conduction electrons. However, they can be present at the surfaces of \cuo{}. 
 Fig.~\ref{fig:cls} shows simulated core level shifts (CLS)  
 for all Cu atoms present in our interface models. In all 
 cases, there are small shifts at the interfaces, that are  much  smaller than the expected CLS between different degrees of oxidation. Hence, 
 only Cu$^{+}$ are present in our interfaces. The CLS have been simulated as the variations in the  electrostatic potential energy averaged inside small spheres 1.94~\AA{} in radius centered at each nucleus. 
Therefore, the existence of our interfaces in CuO$_x$/perovskite solar cells cannot be assured, but 
we hope they can be part of the diversity of this system.

In other group of solar cells, \cuo{} nanocubes have been prepared in solution and deposited upon perovskite surfaces. The cubical shape implies (001) surfaces. 
\citet{cu2o_Liu_advsci2018} achieved a high 18.9~\% efficiency using \cuo{} nanocubes deposited from solution on a mixed perovskite  Cs$_{0.05}$FA$_{0.81}$MA$_{0.14}$PbI$_{2.55}$Br$_{0.45}$. These nanocube surfaces were modified with silane coupling agents, in principle intended to make the nanocrystals soluble. The role of silane coupling in the hole extraction is not totally clear, but comparing the effect  ethenyltriethyloxysilane and octadecyltriethoxysilane, the better performance of the former one was attributed to its shorter tail favoring the hole transfer. 
\citet{cu2o_elseman2019} also obtained efficient solar cells (PCE=17.4~\%) with this method. 
They obtained \cuo{} nanocubes without surfactants and deposited them on MAPI. 
This is probably the system most related with our interface models. 
Their XPS spectra show presence of Cu(II), probably at the nanocube surfaces. 

\citet{cu2o_nejand2016} grew \cuo{} directly over MAPI.  
In that work, prior to \cuo{} growth, the perovskite surface was treated with a  solution of MAI in 2-propanol. This had the main effect of controlling the surface smoothness, 
and possibly also  favored a MAI-terminated surface. For this system, our Cu/MAI and O/MAI models could be 
relevant. 

Let us discuss which of the interfaces studied here would be better for PSC. Optimal band alignment 
occurs when the valence band offset $(\Delta E_V)_{int}$ is zero. Positive band offset causes decrease
of the open circuit voltage, and can also decrease the hole current associated to a non null 
quantum probability of reflection at the potential step. On the other hand, a negative band offset will 
decrease the hole current, but increase the open circuit voltage. 
Device simulations with the wxAMPS and SCAPS softwares  show that small negative values $(\Delta E_{V})_{int} >-0.3$~eV do not affect substantially the solar cell performance, but positive offsets
always decrease the PCE\cite{WangSST2015,haiderJPCS2020}. Attending to these results, 
the O/PbI and Cu/PbI could be the best for solar cell performance. However, as shown in Fig.~\ref{fig:LDOS} 
, there are interface states that could trap conduction electrons, and become recombination centers. 
This could explain why a PCE higher than 20\% has not been reached. In fact, the idea of avoiding 
traps at the \cuo/MAPI interface has been proved experimentally, by means of 
creating    
a buffer layer of spiro-MeOTAD between MAPI and \cuo{}\cite{ChenSciRep2018,cu2o_chen_JMCC2018}. They used the p-i-n inverted structure, and 
the best PCE reached  over 17~\%. 
\citet{Han2018} used the same technique, with the n-i-p structure, proposing that 
buffer spiro-MeOTAD protect the MAPI film from the sputtered nitrogen-doped \cuo. Whatever the 
mechanism is, they improved over the previous n-i-p cells\cite{cu2o_nejand2016}.

Let us consider the effects of the small, but not negligible lattice mismatch between \cuo{} and MAPI, 
which is 2\% with our computational setup. 
For sufficiently thick films, the strain energy can surpass the energy gain associated with the energy of adhesion. The strain energy can be released by means of dislocations at the interface (non-epitaxial growth) or inside MAPI or \cuo{}. In practice, the \cuo{} layer thickness has been treated as an optimization parameter, with values ranging between 4~nm\cite{cu2o_sun2016} and 100~nm\cite{cu2o_nejand2016}. 
The lattice mismatch can be reduced by introduction of bromine in the perovskite. Compositions MAPb(I$_{1-x}$Br$_x$)$_3$, with $x\sim 0.2$ have shorter lattice parameter, and have been  predicted\cite{Jong2016} and verified to improve solar cell stability\cite{jeon2014BrI,zarickreview2018}.
Therefore, the combination of \cuo{} as HTL with MAPb(I$_{1-x}$Br$_x$)$_3$ as photon absorber may be a fruitful path to improve PSC performance. 

Let us consider the implications of MA$^{+}$ dissociation in Cu/MAI and O/MAI interfaces.  
The isoelectric point of \cuo{} has been determined as 7.5 for nanoparticles\cite{iep_salek2015}, 
and  10.2 for commercial samples\cite{iep_wang2019}. 
Regardless of the applicable value, it reveals the fundamentally basic nature of the Cu$_2$O surface,  facilitating deprotonation of MA$^{+}$  by the oxygens at the interface. 
Therefore, preparation of this kind of interface requires a previously deprotonated \cuo{} surface.
If aqueous solution methods are used, this requires pH higher than the isoelectric point. 
Therefore, our interface models are relevant for cases when copper oxide is grown in basic pH, such as in Ref.~\citenum{cu2o_miao_scripta2019}. 
As mentioned above, the Cu/MAI interface allows an energy gain of 0.5~eV per OH formation. This energy represents 0.2~J/m$^2$ in the adhesion energy, which is significant, but not 
enough to destabilize the interface. 
This dissociation of surface methylammonium cations has been associated with fast degradation of MAPI deposited on ZnO\cite{mapionzno} and Al$_2$O$_3$\cite{akbari2017}. 
This fact anticipates a disadvantage for solar cells that include this type of interface, regardless of the alignment of their energy levels. However, methylammonium dissociation at the interface is not sufficient  to degrade MAPI, in absence of a mass transport mechanism. In fact, the fast degradation of MAPI on 
ZnO and Al$_2$O$_3$ has not been observed for \cuo{} and CuO$_x$. 

\section{Conclusions}
\label{sec:conclusions}

Let us summarize our findings.  
 Four atomic scale models have been obtained for the \cuo{}/MAPI interfaces.  The atomic coordinates, 
 electronic states, and band alignments have been obtained by means of first-principles calculations. The 
 interface models differ in the ending atomic layers that join the two materials, which 
 could be controlled with the growth conditions. The formation of copper and oxygen vacancies at the 
 interface layer allows to avoid dangling bonds and deep in-gap electronic states, which is essential for 
 applications in perovskite solar cells. The calculated band alignment is favorable for the photovoltaic 
 conversion using \cuo{} as hole transport layer. 

According to our computed band alignment, and considering device simulations, 
the O/PbI and Cu/PbI interfaces present the best valence band alignments for the solar cells. However, there are interface states with energy close to
 the conduction band minimum of both materials, which could trap electrons and become recombination 
centers. Modification of the interface composition or the formation of a buffer layer is a 
promising pathway to improve solar cell performance.


\begin{acknowledgement}
This work was supported by the CONICYT/FONDECYT Regular 1171807 and Iniciaci\'on 11180984 grants.
Powered@NLHPC: This research was partially supported by the supercomputing infrastructure of the NLHPC (ECM-02).  
Computer resources, technical expertise and assistance 
provided by the Madrid Supercomputing and Visualization Center (CeSViMa) area also acknowledged. This work was partially supported by the Ministerio de Econom\'ia y Competitividad through the project SEHTOP-QC (ENE2016-77798-C4-4-R). S. Botti and T. Rauch are acknowledged for sharing the utility to plot the LDOS.
\end{acknowledgement}

\begin{suppinfo}
Atomic coordinates of slab models, plots of wavefunctions, LDOS, charge densities, and additional tables.
\end{suppinfo}


\providecommand{\latin}[1]{#1}
\makeatletter
\providecommand{\doi}
  {\begingroup\let\do\@makeother\dospecials
  \catcode`\{=1 \catcode`\}=2 \doi@aux}
\providecommand{\doi@aux}[1]{\endgroup\texttt{#1}}
\makeatother
\providecommand*\mcitethebibliography{\thebibliography}
\csname @ifundefined\endcsname{endmcitethebibliography}
  {\let\endmcitethebibliography\endthebibliography}{}

\end{document}